\documentclass[10pt,superscriptaddress,nofootinbib,notitlepage,twocolumn,pra]{revtex4-2}

\usepackage{amsthm,amsmath,amssymb}
\usepackage{mathrsfs}
\usepackage{appendix}
\usepackage{amstext}
\usepackage{graphicx}
\usepackage{url}
\usepackage{float}
\usepackage{bm}
\usepackage{ifthen}
\usepackage[usenames,dvipsnames]{color}
\usepackage{mathrsfs}
\usepackage[colorlinks=true, citecolor=blue, urlcolor=black]{hyperref}
\usepackage{dcolumn}
\usepackage{natbib}
\usepackage{booktabs}
\usepackage{hyperref}
\usepackage{color}
\usepackage{multirow}
\usepackage{mathtools}
\usepackage{lipsum}

\newcommand{\ket}[1]{\mbox{$\left| #1 \right\rangle$}}

\begin{document}

\title{All-Photonic Quantum Repeater for Multipartite Entanglement Generation}

\author{Chen-Long Li}\thanks{These authors contributed equally to this work}
	\affiliation{National Laboratory of Solid State Microstructures and School of Physics, Collaborative Innovation Center of Advanced Microstrucstures, Nanjing University, Nanjing 210093, China}
	\author{Yao Fu}\thanks{These authors contributed equally to this work}	
	\affiliation{Beijing National Laboratory for Condensed Matter Physics and Institute of Physics, Chinese Academy of Sciences, Beijing 100190, China}
	\author{Wen-Bo Liu}
	\author{Yuan-Mei Xie}
	\author{Bing-Hong Li}
	\author{Min-Gang Zhou}
	\affiliation{National Laboratory of Solid State Microstructures and School of Physics, Collaborative Innovation Center of Advanced Microstrucstures, Nanjing University, Nanjing 210093, China}
	\author{Hua-Lei Yin}\email{hlyin@nju.edu.cn}
	\author{Zeng-Bing Chen}\email{zbchen@nju.edu.cn}
	\affiliation{National Laboratory of Solid State Microstructures and School of Physics, Collaborative Innovation Center of Advanced Microstrucstures, Nanjing University, Nanjing 210093, China}




\begin{abstract}
    Quantum network applications like distributed quantum computing and quantum secret sharing present a promising future network equipped with quantum resources.
	Entanglement generation and distribution over long distances is critical and unavoidable to utilize quantum technology in a fully-connected network.
	The distribution of bipartite entanglement over long distances has seen some progresses, while the distribution of multipartite entanglement over long distances remains unsolved.
	Here we report a two-dimensional quantum repeater protocol for the generation of multipartite entanglement over long distances with all-photonic framework to fill this gap.
	The yield of the proposed protocol shows long transmission distance under various numbers of network users.
	With the improved efficiency and flexibility of extending the number of users, we anticipate that our protocol can work as a significant building block for quantum networks in the future.
\end{abstract}

\maketitle

	Quantum entanglement is the central resource to building a quantum network with the applications in computing~\cite{arute2019quantum,zhong2020quantum,zhong2021phase,wu2021strong,broadbent2009universal,stefanie2012demonstration,buhrman2003distributed,zhou2022experimental} and communication~\cite{gisin2002quantum,bennett1992communications,Gu2021differential,chen2007multi,Cao2021coherent,cao2021high}.
	To take advantage of quantum entanglement on a quantum network, one way is generating and directly distributing entangled particles to each user on the network~\cite{pompili2021realization,wengerowsky2018entanglement,siddarth2020trusted,wen2022realizing}.
	Subsequently the users can manipulate the received quantum state by simply measuring the state in a basis to generate correlated classical bits.
	Alternatively, entanglement can be realized by a postselected measurement-device-independent (MDI) method~\cite{lo2012measurement} to establish classical correlation~\cite{fu2015long} without any entanglement source.
	In this scenario, each user on the network prepares and transmits quantum states to an untrusted relay.
	The untrusted relay performs Bell state measurement or Greenberger-Horne-Zeilinger (GHZ) state measurement in bipartite or multipartite cases respectively and publicizes whether the measurement succeeds.
	Given that each user prepares the states in the same basis, a successful measurement implies successful preparation of postselected Bell or GHZ states.
	The postselecting measurement device held by the untrusted relay can be regarded as a black box and such scheme is immune to all detection-side attacks in cryptographic tasks.

	In either case, photons are adopted as information carrier in general and are transferred through optical fiber channel.
	As a result, the transmission efficiency is limited by channel loss, which further limits the deployment of a large-scale quantum network.
	Hence, quantum repeaters are needed to overcome the limitations and improve the transmission efficiency. 
	In memory-based quantum repeater protocols~\cite{duan2001long, simon2007quantum,fault2006childress, hybrid2006loock}, quantum memories are necessary to be entangled with photons and preserve entanglement at least until receiving heralding signals of successful entanglement swapping. 
	In this process, time multiplexing from quantum memories’ preserving entanglement enhances transmission efficiency.
	In all-photonic quantum repeater~\cite{azuma2015all}, cluster states are utilized to demonstrate polynomial scaling of transmission efficiency with distance.
	Specifically, each user prepares and sends multiple single photons maximally entangled with local qubits to adjacent receiver node in all-photonic quantum repeater.
	At the same time, any other source node prepares the encoded complete-like cluster state and sends the left-arms (right-arms) to the left-hand (right-hand) adjacent receiver node.
	The receiver node applies Bell state measurement to the 2nd-leaf of received qubits.
	Since the Bell state measurement succeeds probabilistically, the receiver node performs the $X$- or $Z$-basis measurement adaptively and the bipartite entanglement can be established.
	The aforementioned process in fact uses the idea of spatial multiplexing and adaptive operation to avoid the usage of matter quantum memories.
	Similarly, the idea of spatial multiplexing is applied in adaptive MDI quantum key distribution protocol~\cite{azuma2015allqkd}.
	In this scheme, both users send multiple single photon states simultaneously to the central relay who subsequently confirms the arrival of photons by applying quantum non-demolition (QND) and pairs the arrived photons adaptively.
	
	Currently, quantum memories are still challenging in experiment and far from practical use.
	Furthermore, most quantum repeater protocols are one-dimensional and distribute pair-wise entanglement.
	However, network applications involving more than two users, such as anonymous transmission~\cite{christ2005quantumanonymous}, secret sharing~\cite{hillery1999qss}, leader election~\cite{andris2004multiparty}, clock stabilization~\cite{komar2014quantum}, and conference key agreement~\cite{chen2007multi} require the establishment of genuine multipartite entangled states.
	In this work, we propose a two-dimensional quantum repeater for the generation of multipartite entanglement---typically, the GHZ entangled states---over long distances on the quantum network with all-photonic framework.
    We utilize the principle of spatial multiplexing to establish multipartite entanglement with improved efficiency based on the result in all-photonic quantum repeater and adaptive MDI quantum key distribution.
    Specifically, spatial multiplexing means simultaneous transmission of multiple quantum signals using multiplexing as mentioned before.
    The yield of the proposed two-dimensional quantum repeater investigated under state-of-the-art experimental parameters shows the scalability and flexibility of our quantum repeater protocol.
	We believe our protocol manifests potential to be an indispensable building block for quantum networks both theoretically and technologically.

	\begin{figure}[tbp]
		\includegraphics[width=8cm]{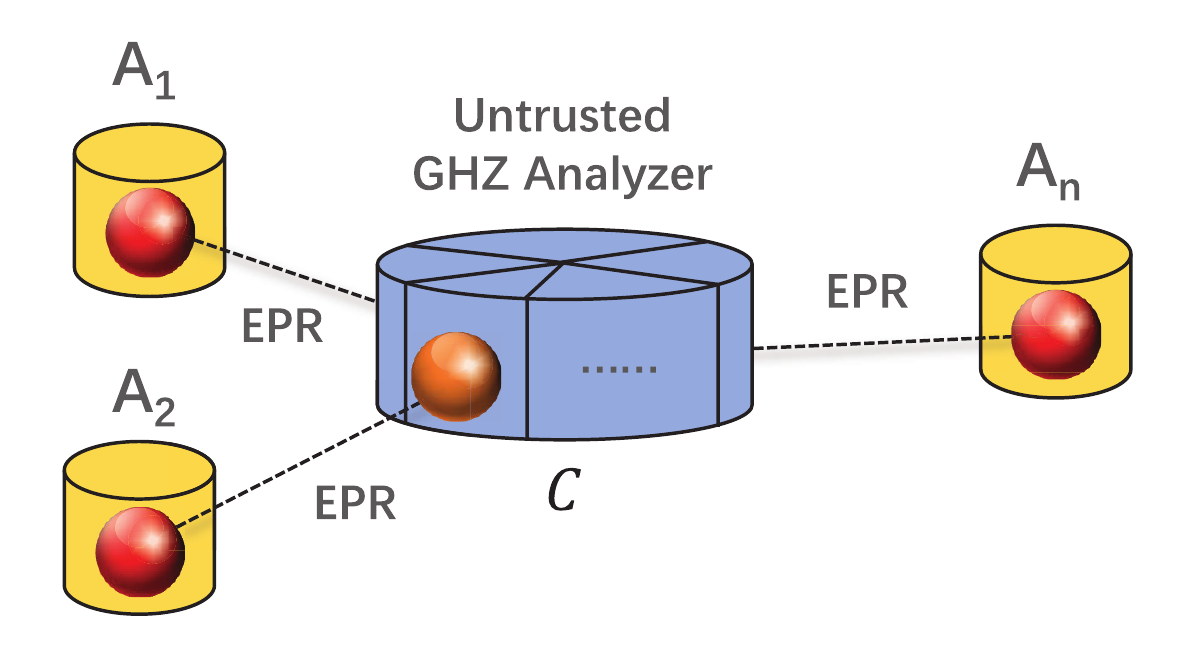}
		\caption{Schematic of the proposed two-dimensional quantum repeater for generation of multipartite entanglement over long distances. In this scheme, $(A_1,...,A_{n})$ transmit particles belonging to one of the Bell entanglement with spatial multiplexing to the central node $C$. The central node $C$ performs entanglement swapping adaptively.  
		}\label{figrep}
	\end{figure}
	
	In the following we present the protocol of establishing multipartite entanglement as shown in Fig.~\ref{figrep}.
	Here we denote the $i$th user and the untrusted GHZ analyzer as $A_i$ $(i=1,...,n)$ and GHZ analyzer as node $C$, respectively.
	
	\begin{itemize}
	    \item[$(i)$] 
	    $A_i$ generates $M$ Bell pairs $\ket{HH}+\ket{VV})/\sqrt{2}$.
	    Then $A_i$ keeps one of the two particles in the Bell entanglement pair and transmits the other to node $C$ simultaneously using spatial multiplexing.
	    
	    \item[$(ii)$]
	    Node $C$ performs QND measurements to confirm the arrival of single photons from $(A_1,...,A_{n})$.
	    
	    \item[$(iii)$]
	    After the QND measurements, the confirmed photons from every user form a group via optical switches. Node $C$ then performs GHZ projection on the group to realize entanglement swapping. 
	    Note that each party should successfully transmit at least one single photon through QND measurements. 
	    Otherwise, this trial is considered to be failed.
	    
	    \item[$(iv)$]
	    Node $C$ announces the group information and the result of entanglement swapping.
	    Each $A_i$ keeps the information of states that were successfully projected onto the GHZ state and discards the rest.
	    Furthermore, nearly perfect GHZ entanglement can be established between $n$ users by entanglement purification.
	\end{itemize}

	\begin{figure}[tp!]
		\includegraphics[width=7.5cm]{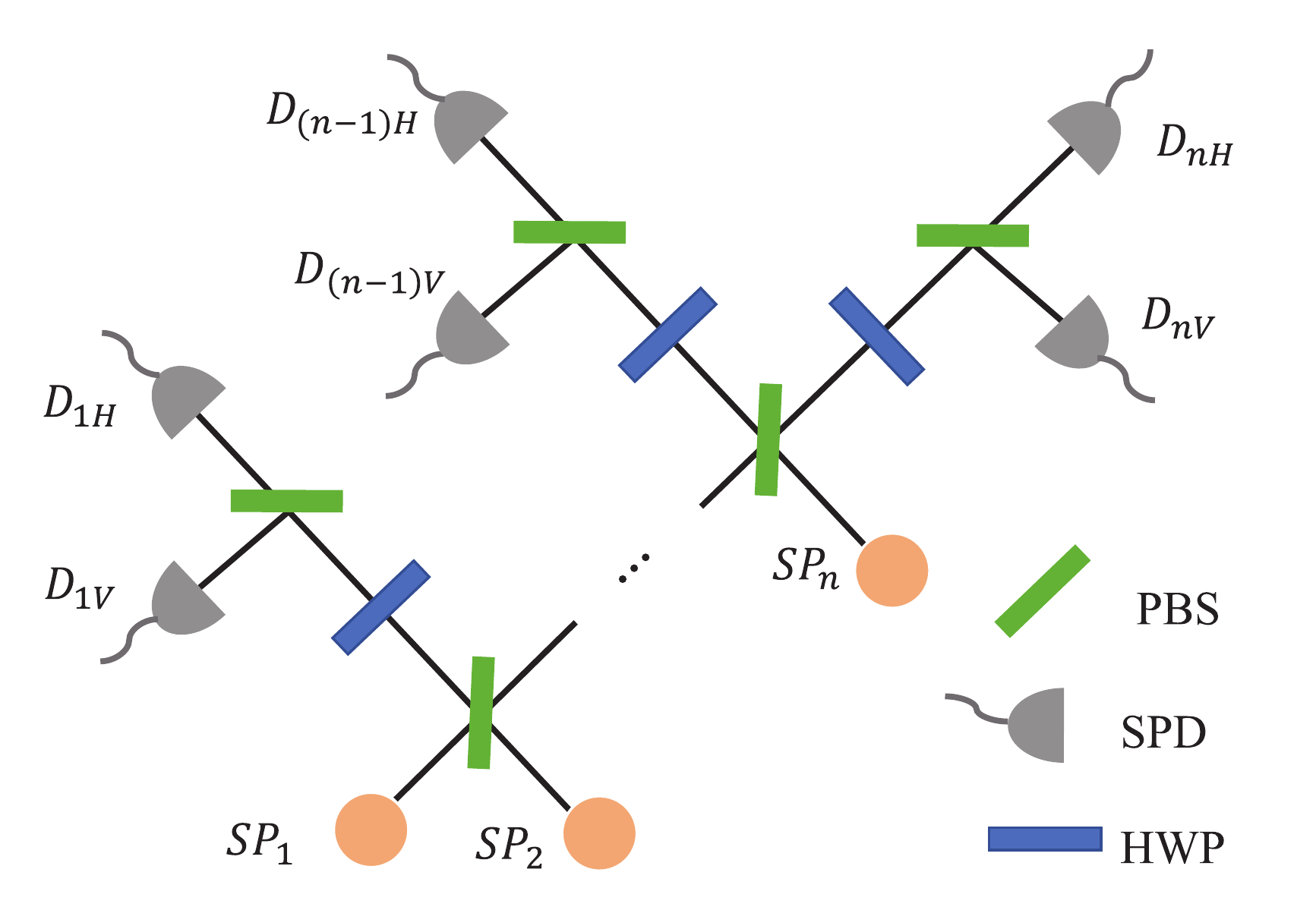}
		\caption{Schematic of Greenberger-Horne-Zeilinger (GHZ) analyzer based on linear optical elements. $\{A_i\}_{i=1,...,n}$: input modes; PBS: polarizing beam splitter which transmits $\ket{H}$ and reflects $\ket{V}$ polarizations; HWP: half-wave plate used to conduct a $45^{\circ}$ rotation of polarization. SPD: single photon detector denoted as $\{D_{iH},D_{iV}\}_{i=1,...,n}$ in the figure.    
		}\label{fig3}
	\end{figure}
	
	We consider the performance of the protocol in constructing classical correlated bits at long distances by measuring the remaining particles at each user after entanglement purification.
	Specifically, each user randomly chooses a basis from $X$ and $Z$ and measure the remaining particles.
	Therefore, each user holds a string of correlated classical bits.
	After entanglement purification, the GHZ states shared among the $n$ users are nearly unentangled with outer malicious party due to the monogamy of entanglements, which means the information leakage is negligible.
	Therefore, the classical correlated bits are information-theoretically secure, which can be used for cryptographic task like group encryption and decryption.
	
	Maneva and Smolin proposed a multiparty hashing protocol to distill nearly perfect GHZ states~\cite{maneva2002improved}.
	The yield per input mixed state of this protocol when $n=3$ is 
	\begin{equation}
	    D=1-\max\{h(e_{b_1}),h(e_{b_2})\}-h(e_p).
	\end{equation}
	Here, $h(x)=-x\log_2x-(1-x)\log_2(1-x)$ is the standard binary Shannon entropy function.
	$e_{b_1}$ and $e_{b_2}$ are marginal bit error rates and $e_p$ is the phase error rate.
	Taking the transmission of photons in a lossy channel into account, the yield of our protocol with $n$ users in the case of asymptotic data can be generalized easily and given by
	\begin{equation}
	    D=Q\left[1-\max\{h(e_{b_1}),...,h(e_{b_n})\}-h(e_p)\right],
	\end{equation}
	where $Q$ is the probability of a successful GHZ projection.
	
	The success probability $Q$ is defined as efficiency of generating groups of successfully transmitted photons.
	Specifically, we have 
	\begin{equation}\label{suc_prob}
		Q=\frac{\bar{N}}{M},
	\end{equation}
	where $\bar{N}$ is the average number of groups formed by successfully transmitted photons using $M$ multiplexing.
	Intuitively, if we denote transmission efficiency of the channel from any $i$th user to the GHZ analyzer as $\eta$, and $M$ multiplexing is used, then $\bar{N}\sim M\eta$.
	Therefore, we have $Q\sim\eta$.
	The approximate relation can be converted to an equation $Q=\eta$ under the asymptotic limit ($M\rightarrow\infty$). 
	We prove this equation when $n=3$ in Supplement 1.
	To guarantee that more than on average one group is generated before entering the GHZ analyzer, the number of multiplexing should satisfy $M\ge \eta^{-1}$, which is actually implies that $\bar{N}\sim M\eta\ge1$.
	
	\begin{table}[tp!]
		\caption{Different clicks to identify $\ket{\Phi^{+}_0}$ and $\ket{\Phi^{-}_0}$. In this table, we show the corresponding clicks on $V$ to identify $\ket{\Phi^{+}_0}$ and $\ket{\Phi^{-}_0}$ when $n$ is odd and even.}
		\begin{tabular}{ccc}
			\hline
			&$n$ is odd&$n$ is even\\
			\hline
			$\ket{\Phi^{+}_0}$&even number of clicks&odd number of clicks\\
			$\ket{\Phi^{-}_0}$&odd number of clicks&even number of clicks\\
			\hline
		\end{tabular}	\label{tab1}
	\end{table}
	
	In our simulation, we consider the GHZ analyzer based on linear optical elements~\cite{pan1998greenberger} as shown in Fig.~\ref{fig3}, which is composed of just polarizing beam splitters (PBS) and half-wave plates (HWP).
	This GHZ analyzer is capable of identifying two of the $n$-particle GHZ states.
	To be specific, by considering the evolution of $n$-particle GHZ state $\ket{\Phi^{\pm}_0}=1/\sqrt{2}(\ket{HHH...H}\pm\ket{VVV...V})$ in the GHZ analyzer, we can identify $\ket{\Phi^{+}_0}$ and $\ket{\Phi^{-}_0}$ from different clicks.
	We summarize the result of different clicks and corresponding state in Table~\ref{tab1} for easier reference.
	
	Photons travel through optical fiber channels whose transmittance is determined by $\eta_{\text{channel}}=\exp\left(-\frac{l}{l_{\text{att}}} \right)$, where the attenuation distance $l_{\text{att}}=27.14$ km and $l$ is the distance from any $i$th user to the GHZ analyzer.
	QND measurements are required to confirm the arrival of photons and the success probability of QND measurements is denoted by $p_{\text{QND}}$.
	To simplify the simulation, we consider a QND measurement for a single photon based on quantum teleportation~\cite{kok2002single} with ideal parameters where we have $p_{\text{QND}}=1/2$. 
	Active feedforward technique is needed to direct the arrived photons to the GHZ analyzer via optical switches.
	We assume the active feedforward costs time $\tau_a=67$ ns~\cite{ma2011experimental}, which is equivalent to a lossy channel with the transmittance $\eta_a=\exp(-\tau_ac/l_{\text{att}})$, where $c=2.0\times10^8$ $\text{ms}^{-1}$ is the speed of light in an optical fiber.
	Single photon detectors in the GHZ analyzer are characterized by an efficiency of $\eta_d=0.93$ and a dark count rate of $p_d=1\times10^{-9}$~\cite{minder2019experimental}, by which we can estimate the success probability of GHZ projection in the $X(Z)$ basis $Q_{X(Z)}^{\text{GHZ}}$.
	Based on the aforementioned assumption on experiment parameters, we analytically estimate the gain with
	\begin{equation}
		Q=Q^{\text{GHZ}}_{X(Z)}\cdot p_{\text{QND}}\cdot \eta_{\text{channel}}\cdot \eta_a.
	\end{equation}
	See Supplement 1 for the concrete process of estimation of the marginal bit error rates and phase error rate. 
	
	\begin{figure}[tbp]
		\includegraphics[width=8cm]{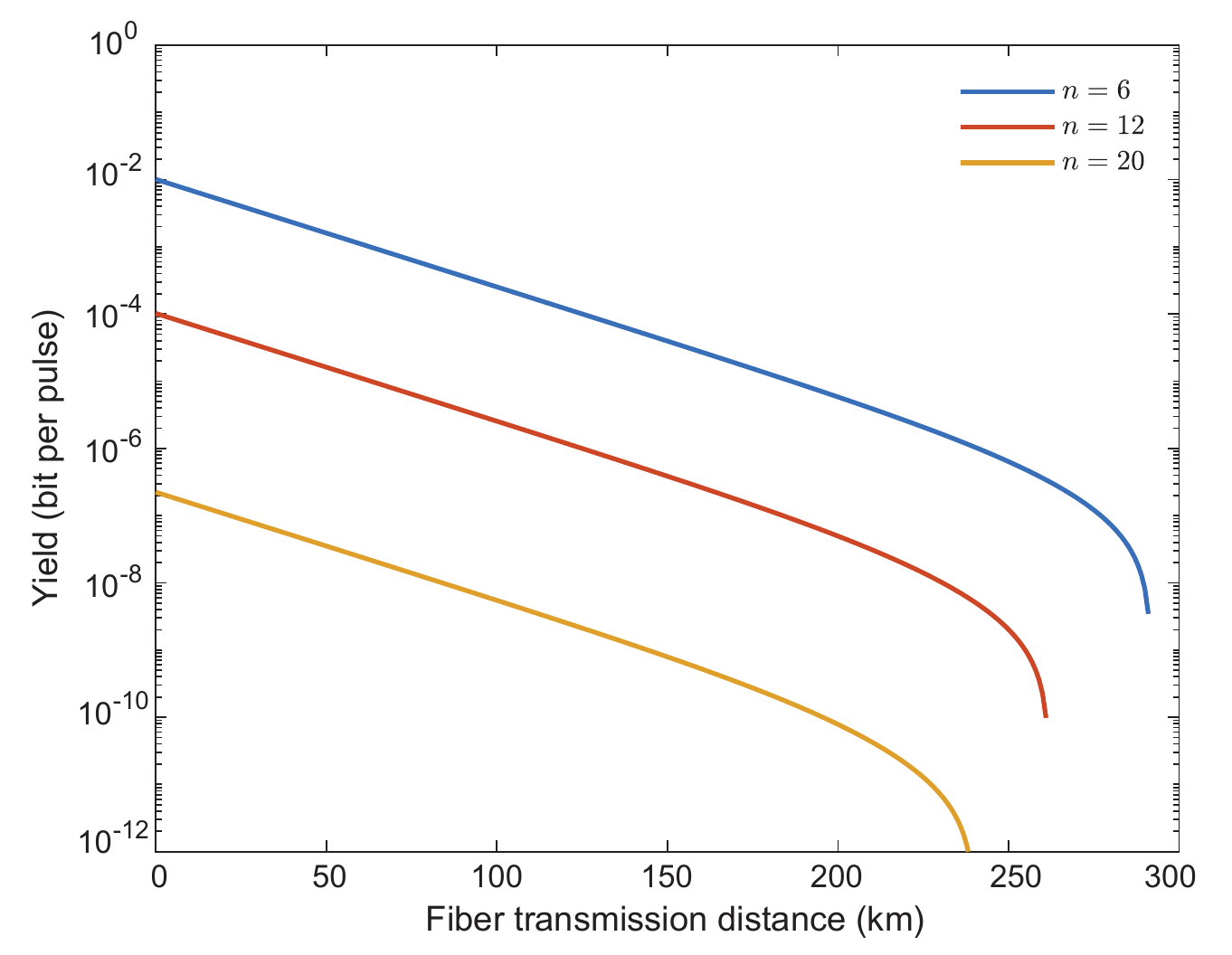}
		\caption{Yield of our two-dimensional quantum repeater framework. We plot the entanglement generation yield of our scheme with experimental parameters discussed in the main text when $n=6,12,20$ in different colors. The fiber transmission distance of horizontal axis denotes the distance between any $i$th user and the untrusted GHZ analyzer.
		}\label{rep_perform}
	\end{figure}
	In Fig.~\ref{rep_perform}, we plot the yield of the proposed quantum repeater framework with the aforementioned experimental parameters under $n=6,12,20$.
	The result shows that the slopes of the yield remained unchanged with different values of $n$, which stems from the spatial multiplexing and adaptive operation used in our protocol.
	The three curves show a transmission distance about 250 km for $n=6,12,20$.
	Therefore, we anticipate our quantum repeater fits well in intra- and intercity network applications equipped with quantum resources.
	
	Our two-dimensional quantum repeater protocol incorporates GHZ projection and spatial multiplexing to directly generate multipartite entanglement over long distances under any number of communication parties.
	Multiparty quantum communication protocols can then be realized without quantum memories due to the all-photonic nature of our scheme.
	Combined with one-dimensional quantum repeaters, we can expect the construction of an inter-continent quantum network in the future.
	
	In this work, we report a two-dimensional quantum repeater protocol for generating GHZ entanglement at long distances with enhanced efficiency by using the principle of spatial multiplexing and adaptive operation. 
	We investigate the yield of the proposed quantum repeater protocol within mild experimental parameters when establishing classical correlations over multiple users.
	The result shows that our protocol has great potential to be deployed over long distances with any number of users.
	Based on the result of this work, we can anticipate a wide and flexible usage of our work in multiparty applications of secure quantum network.
	
	Here we remark on possible directions for future work.
	As we have mentioned before, the all-photonic quantum repeater utilizes cluster states to realize a polynomial scaling with distance which is in fact a result of spatial multiplexing. 
	Therefore, with such spatial multiplexing idea, we can develop other quantum communication protocol with enhanced efficiency which is a polynomial scaling with distance under rather mild experiment requirement.
	Further study can also be conducted on evaluating the performance of our protocol with complete GHZ analyzer and reducing the need of large-scale optical switches which will cause unwanted loss in protocol.

\section*{Acknowledgements}
We gratefully acknowledge the supports from the National Natural Science Foundation of China (No. 12274223), the Natural Science Foundation of Jiangsu Province (No. BK20211145), the Fundamental Research Funds for the Central Universities (No. 020414380182), the Key Research and Development Program of Nanjing Jiangbei New Aera (No. ZDYD20210101), and the Program for Innovative Talents and Entrepreneurs in Jiangsu (JSSCRC2021484). 


%

\end{document}